\pdfoutput=1 
\documentclass[useAMS,usenatbib]{mn2e}
\usepackage{aas_macros}
\usepackage[]{graphicx}
\usepackage{amssymb}
\title[Cusp-core transformations in dwarf galaxies]{Cusp-core transformations in dwarf galaxies: observational predictions} 
\author[R. Teyssier et al.]{\parbox[t]{\textwidth}{Romain Teyssier$^{1,4}$\thanks{E-mail: romain.teyssier@gmail.com}, 
Andrew Pontzen$^{2}$, Yohan Dubois$^{3}$ and Justin I. Read$^{5,6}$ }\vspace*{6pt}\\
$^{1}$Institute for Theoretical Physics, University of Zurich, CH-8057 Z\"urich, Switzerland\\
$^{2}$Astrophysics, University of Oxford, Denys Wilkinson Building, Keble Road, OX13RH Oxford, UK\\
$^{3}$Institut d'Astrophysique de Paris, 98bis boulevard Arago, 75014 Paris, France\\
$^{4}$CEA Saclay, DSM/IRFU/SAP, B\^atiment 709, F-91191 Gif-sur-Yvette, Cedex, France\\
$^{5}$Department of Physics and Astronomy, University of Leicester, University Road, LE17RH Leicester, UK\\
$^{6}$Institute for Astronomy, Department of Physics, ETH Z\"urich, Wolfgang-Pauli-Strasse 27, CH-8093 Z\"urich, Switzerland}

\begin{document}

\maketitle

\label{firstpage}

\begin{abstract}
The presence of a dark matter core in the central kiloparsec of many dwarf galaxies has been a long standing
problem in galaxy formation theories based on the standard cold dark matter paradigm. Recent  simulations, based on Smooth Particle Hydrodynamics and rather strong feedback recipes have shown that it was indeed possible to form extended dark matter cores using baryonic processes related to a more realistic treatment of the interstellar medium. Using adaptive mesh refinement, together with a new, stronger supernovae feedback scheme that we have recently implemented in the RAMSES code, we show that it is also possible to form a prominent dark matter core within the well-controlled framework of an isolated, initially cuspy, 10 billion solar masses dark matter halo. Although our numerical experiment is idealized, it allows a clean and unambiguous  identification of the dark matter core formation process. Our dark matter inner profile is well fitted by a pseudo-isothermal profile with a core radius of 800~pc. The core formation mechanism is consistent with the one proposed recently by Pontzen \& Governato. We highlight two key observational predictions of all simulations that find cusp-core transformations: (i) a bursty star formation history (SFH) with peak to trough ratio of 5 to 10 and a duty cycle comparable to the local dynamical time; and (ii) a stellar distribution that is hot with $v/\sigma \sim 1$. We compare the observational properties of our model galaxy with recent measurements of the isolated dwarf WLM. We show that the spatial and kinematical distribution of stars and HI gas are in striking agreement with observations, supporting the fundamental role played by stellar feedback in shaping both the stellar and dark matter distribution. 
\end{abstract}

\begin{keywords}
galaxies: formation, dwarf -- cosmology: dark matter -- methods: numerical
\end{keywords}

\section{Introduction}

Dwarf galaxies, although very numerous and common in our present day universe, are also very faint and difficult to observe. 
Nevertheless, it is now established that star formation proceeds at a very inefficient rate in dwarf galaxies, making them ideal laboratories to study the spatial distribution of their parent dark matter halo. Indeed, if dwarf galaxy are dark matter dominated, 
a stellar kinematic analysis gives direct constraints on the dark matter mass distribution. Although theoretical predictions of pure N-body models favor the formation of a cusp in the inner region of dark matter halos \citep{1994Natur.370..629M, 1997ApJ...490..493N}, the observed rotation curve of dwarf and low surface brightness galaxies was shown to be more consistent with a shallower profile, even a constant density core \citep{1997MNRAS.290..533D,2001ApJ...552L..23D,2002A&A...385..816D,2008ApJ...676..920K}. This apparent disagreement led to the so-called ``cusp-core'' problem, a serious challenge for the currently favored Cold Dark Matter (CDM) paradigm \citep{1994Natur.370..629M,1994ApJ...427L...1F}. Many solutions to this problem have been proposed in the recent years -- for example a warm dark matter particle \citep{2010ApJ...710L.161K,2011JCAP...03..024V,2012arXiv1202.1282M} or self-interacting dark matter \citep{2000PhRvL..84.3760S, 2012arXiv1208.3025R} -- but the only explanation consistent with CDM relies on the effect of baryonic physics, and more precisely on stellar feedback, to modify substantially the dark matter distribution.

\cite{2008Sci...319..174M} were the first to find a cusp-core transformation effect in a cosmological simulation, modeling a dwarf galaxy of mass $\sim 10^9$\,M$_\odot$ down to redshift $z \sim 5$. This has been seen again in more recent work at higher masses and reaching lower redshifts \citep{2010Natur.463..203G, 2012arXiv1202.1282M,2012MNRAS.422.3081M}. A long history of work has looked at the possibility of baryonic physics generating such a transformation: for instance  \cite{1996MNRAS.283L..72N} showed that impulsive mass loss leads to irreversible expansion of orbits near the centre,  \cite{2005MNRAS.356..107R}  suggested that repeated epochs of outflows could produce a strong enough effect to generate cores and \cite{2006Natur.442..539M} pointed out that internal bulk motions of gas (not necessarily outflows) could have the same effect. Finally, \cite{2012MNRAS.421.3464P} (hereafter PG12)  produced an analytic model of impulsive heating which was able to make quantitative predictions for the rate of cusp flattening, agreeing with the simulated results of \cite{2010Natur.463..203G}. This demonstrated that impulsive gas motions were the dominant cause of the simulated cusp-core transformations as opposed to, for example, heating due to dynamical friction \citep[e.g.][]{2001ApJ...560..636E,2010ApJ...725.1707G}.

Although cosmological simulations provide realistic environments and mass accretion histories for the model galaxies,
they are often challenging to analyze because of their geometrical and evolutionary complexity. Moreover, since they are  time consuming, the force resolution is currently limited to $\sim$~80 pc, which is only one tenth of the measured core radius \citep{2010Natur.463..203G}. It is therefore quite important to perform additional simulations with both a simpler set up and a higher resolution, which is precisely the goal of our paper. In \cite{2006Natur.442..539M}, a similar approach was proposed with an isolated NFW halo stirred by 3 arbitrarily moving gas clumps. Although the proposed set-up was quite simplistic, it provided a well controlled experiment of cusp-to-core evolution with a better force resolution of 40 pc. Very recently, \cite{2012MNRAS.423..735C} have performed a series of isolated NFW halo simulations with a more realistic treatment of star formation, stellar feedback and the induced gas motions based on the SPH technique. This work, following up on a series of papers reproducing many properties of the gas and stellar distribution in dwarf galaxies \citep{2008MNRAS.389.1111V, 2011MNRAS.416..601S}, also confirmed the formation of a dark matter core.

In this paper, we will also perform idealized simulations of an isolated NFW halo, with a more realistic treatment of the ISM physics, and, for the first time, with the Adaptive Mesh Refinement (AMR) code RAMSES \citep{2002A&A...385..337T}. It is indeed important to verify that core formation can also be recovered using a different type of code than the one used so far in all the previously cited papers, namely SPH in two different implementations, GADGET by \cite{2005MNRAS.364.1105S} and GASOLINE by \cite{2004NewA....9..137W}. As shown by \cite{2007MNRAS.380..963A} and \cite{2008MNRAS.387..427W}, both code types suffer from different systematic effects that might affect the numerical solution, especially when trying to resolve the clumpy ISM (see also \cite{2010MNRAS.405.1513R} and \cite{2012MNRAS.422.3037R}).  

The paper is organized as follows: in the first section, we describe the set up of our numerical experiment, with 3 different simulations that we would like to compare. We then present and discuss our new stellar feedback implementation in the RAMSES code, which allows us to have a much stronger dynamical effect on the surrounding gas than we were previously able to achieve, without invoking extra sources of energy. In the third section, we present our results, with emphasis put on the dark matter distribution and how the formation of the core correlates with strong potential fluctuations triggered by powerful gas expulsion phases. Finally, we discuss and interpret our results in the light of the recently proposed mechanism by PG12, and confront our findings with observations.

\section{Simulation set-up}

\begin{figure}
    \includegraphics[width=0.5\textwidth]{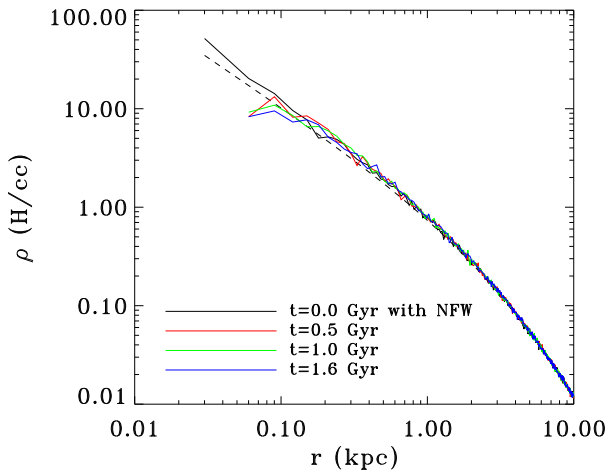}
  \caption{ Evolution of the dark matter density profile over the 2~Gyr of evolution for the control run with adiabatic (no cooling and no star formation) hydrodynamics. We see that the dark matter halo density profile remains stable (with at most 10\% deviations), except within the central 60~pc, which corresponds to twice the cell size in this case, in which we see less than a factor of two deviations from the initial profile.}
  \label{fig:dmorun}
\end{figure}

We consider an isolated halo in hydrostatic equilibrium with $f_{\rm gas}=15~\%$. Both gas and dark matter follows the same NFW density profile, with concentration parameter $c=10$. The halo circular velocity is chosen to be $V_{200}=35$~km/s, corresponding to the virial radius $R_{200}=50$~kpc (or 35~kpc/h) and the virial mass $M_{200}=1.4\times10^{10}~M_{\odot}$ (or $10^{10}~M_{\odot}/h$). The halo is truncated at 112.5 kpc, so that the total enclosed mass is $M_{tot}=2\times 10^{10}~M_{\odot}$.  Following now rather standard prescriptions, such as in \cite{2007MNRAS.375...53K} and \cite{2008A&A...477...79D}, we initialized the gas temperature by solving the hydrostatic equilibrium equation \cite[see Eq. 1 in][for example]{2007MNRAS.375...53K}, and we set the gaseous halo in slow rotation around the z-axis, using the average angular momentum profile computed from cosmological simulation \citep{2001ApJ...555..240B} and a spin parameter $\lambda=0.04$. For more details on the set-up, the reader can refer to \cite{2008A&A...477...79D}. 
We use 1 million dark matter particles to sample the phase space of our dark matter halo. Initial positions and velocities were computed using the density-potential pair approach of \cite{2004ApJ...601...37K}, later refined by \cite{2006MNRAS.367..387R}. To test the stability of our gas-dark matter equilibrium system, we ran a first simulation with only adiabatic gas dynamics, without cooling nor star formation. As can be seen in Figure~\ref{fig:dmorun}, our dark matter halo is stable during 2~Gyr of evolution and does not deviate by more than 10\% from the initial NFW distribution, except within 60~pc from the center (corresponding roughly to twice the cell size), where the deviation can be as high as a factor of 2.

All the simulations presented here have been run using the RAMSES code \citep{2002A&A...385..337T, 2006JCoPh.218...44T, 2006A&A...457..371F}. We used a quasi-Lagrangian refinement strategy to build the initial AMR grid, as well as refining and de-refining cells during the course of the simulation:~each cell is individually refined if it contains more than 8 dark matter particles or if it contains a baryonic mass (gas mass + star particle mass) larger than $8\times m_{res}$, where $m_{res}=1500~M_{\odot}$. Note that our gas mass resolution corresponds initially to 2~million gas resolution elements across the whole gaseous halo. The box size being set to $L=300$~kpc (in our terminology this corresponds to the first AMR level $\ell=1$), we used isolated boundary conditions for the Poisson solver and zero-gradient boundary conditions for the hydro solver. For the latter, we used the HLLC Riemann solver and the MinMod slope limiter (see \cite{2006A&A...457..371F} for related details).

Without cooling and star formation, we do not need to specify a maximum level of resolution: we just let the code refine the grid until it runs out of mass in the smallest AMR cell. In our adiabatic simulation, we have indeed reached ``only'' level $\ell=13$ at the very centre of the halo, corresponding  to a cell size $\Delta x=36$~pc. If cooling and star formation are activated, gas dynamics becomes strongly dissipative and nothing can prevent the collapse. It is therefore crucial to set a maximum level of refinement, or a minimum grid resolution. In this paper, we use $\ell_{\max}=14$ or $\Delta x=18$~pc, which is only a factor of 2 better than in the dark matter only case, therefore avoiding two-body relaxation in the dark matter component. Our limited spatial resolution requires a careful treatment of our thermal model, in order to avoid numerical difficulties. As is now customary in galaxy and star formation simulations, we use an artificial pressure floor designed to enforce that the effective Jeans length is equal to 4 computational cells \citep{1997ApJ...489L.179T}. This extra-pressure, equal to $P_J \simeq 16G\Delta x^2 \rho^2$, is added to the thermal pressure in the Euler equation, gas cooling being applied only to the thermal internal energy. We use standard H and He cooling \citep{1996ApJS..105...19K}, with an additional contribution from metals based on the \cite{1993ApJS...88..253S} model above $10^4$~K and metal fine-structure cooling below $10^4$~K as in \cite{1995ApJ...440..634R}. Metallicity is modeled as a unique, passively advected quantity, noted $Z$, representing the mass fraction of metals, and seeded by individual supernovae event with a yield $y=0.1$. The initial metallicity in the halo is set to $Z=10^{-3}~Z_{\odot}$, enabling from the start cooling down to very low temperatures. With these ingredients and for our spatial resolution of 18~pc, gas will cool efficiently down to $T_J \simeq 600$~K and reach a density of $n_J \simeq 60$~H/cc, before hitting the Jeans-length related pressure floor (the index $J$ standing for Jeans-length related quantities). This is a conservative estimate of the maximum gas density we can reach before being affected by finite resolution effects. 

Star formation (SF) is modeled using a Schmidt law \citep{1959ApJ...129..243S}, for which the SF rate is given by
\begin{equation}
\dot \rho_* = \epsilon_* \frac{\rho_{\rm gas}}{t_{\rm ff}}~~~{\rm if}~~~\rho_{\rm gas} > \rho_*{\rm .}
\end{equation}
The SF efficiency per free-fall time $t_{\rm ff}$ is usually chosen close to a few percent \citep{2007ApJ...654..304K}; we used here $\epsilon_*=0.01$. The density threshold for SF, above which gas is considered to be dense enough to be eligible to form stars, has to be chosen carefully. In order to mimic as closely as possible realistic SF in dense molecular clouds, we want to choose this threshold as high as possible. On the other hand, as discussed previously, the gas density distribution in the simulation will be affected by finite resolution effect close to the Jeans density, in our case $n_J \simeq 60$~H/cc. It is therefore quite natural to choose the SF threshold density close to the Jeans density. In the present paper, we conservatively picked $n_*=n_J/3=20$~H/cc. These recipes have been all implemented in the RAMSES code and used for many galaxy formation studies in the recent years \citep{2009MNRAS.397L..64A, 2010ApJ...720L.149T, 2011ApJ...730....4B, 2011MNRAS.410.1391A, 2012MNRAS.tmp.2970S}. We will now present in more details the main ingredient of the present study, namely our new implementation of stellar feedback.

\section{Stellar feedback implementation}

\begin{figure*}
    \includegraphics[width=0.3\textwidth]{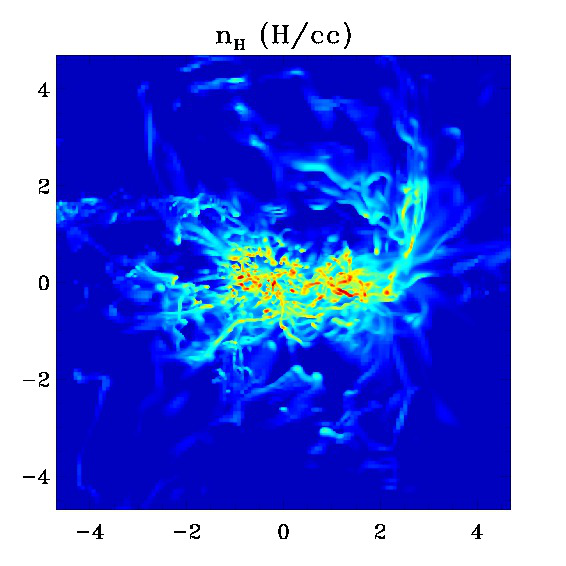}
    \includegraphics[width=0.3\textwidth]{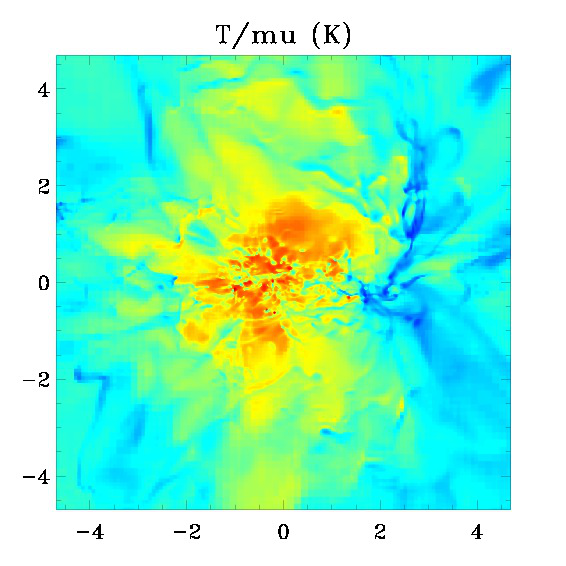}
    \includegraphics[width=0.3\textwidth]{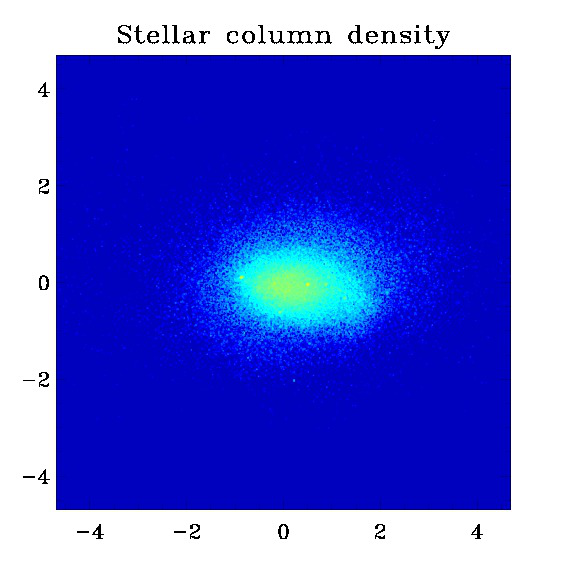}\\
    \includegraphics[width=0.3\textwidth]{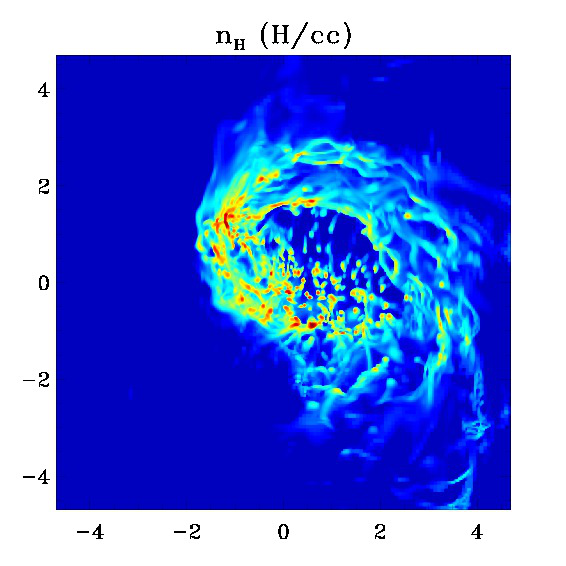}
    \includegraphics[width=0.3\textwidth]{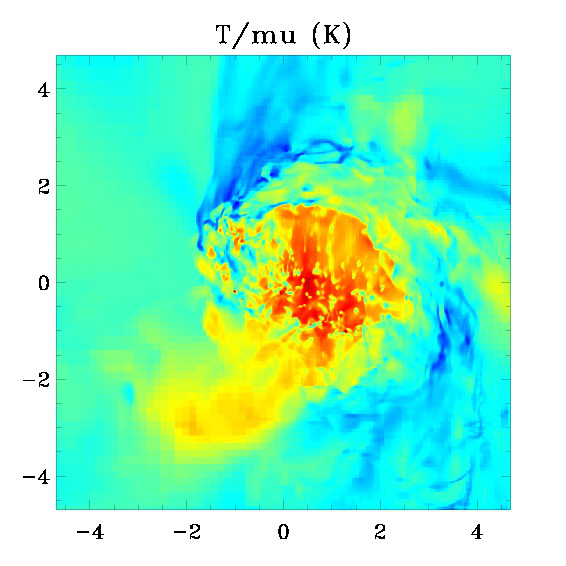}
    \includegraphics[width=0.3\textwidth]{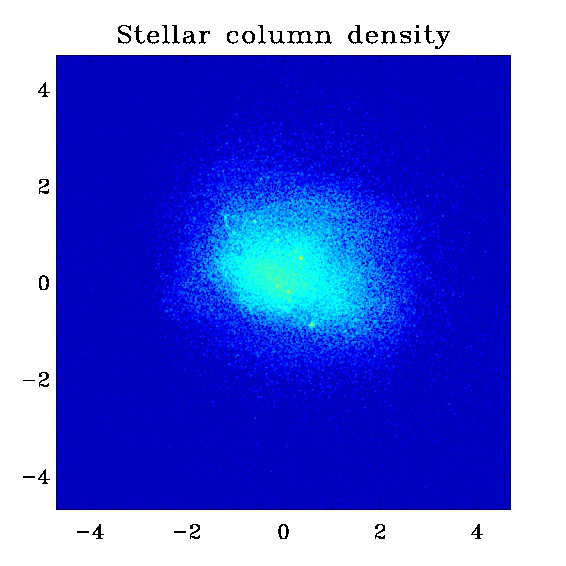}\\
    \includegraphics[width=0.3\textwidth]{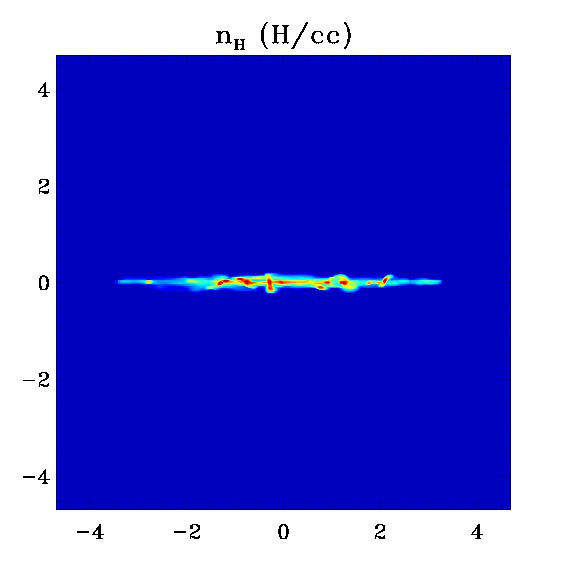}
    \includegraphics[width=0.3\textwidth]{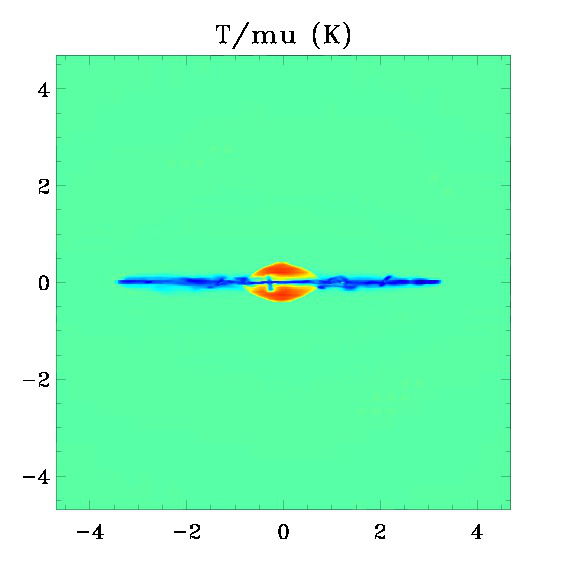}
    \includegraphics[width=0.3\textwidth]{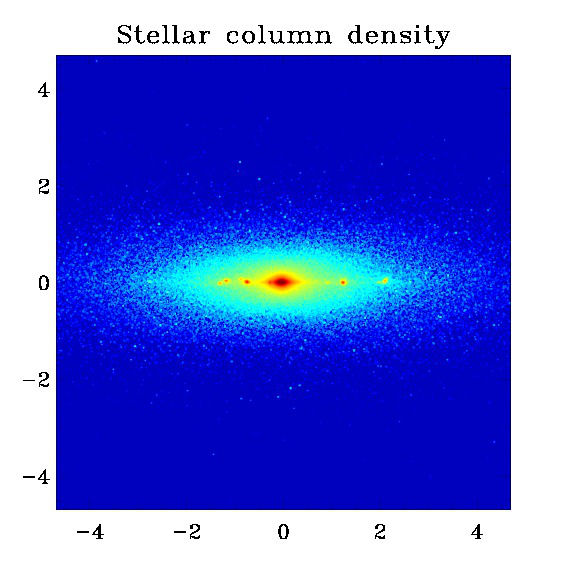}\\
    \includegraphics[width=0.3\textwidth]{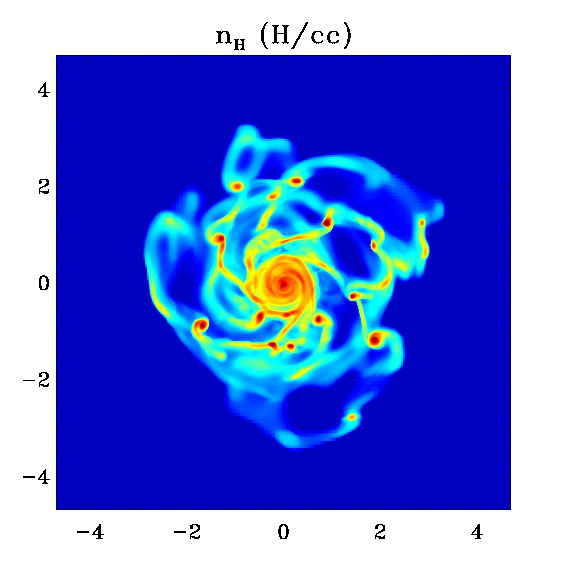}
    \includegraphics[width=0.3\textwidth]{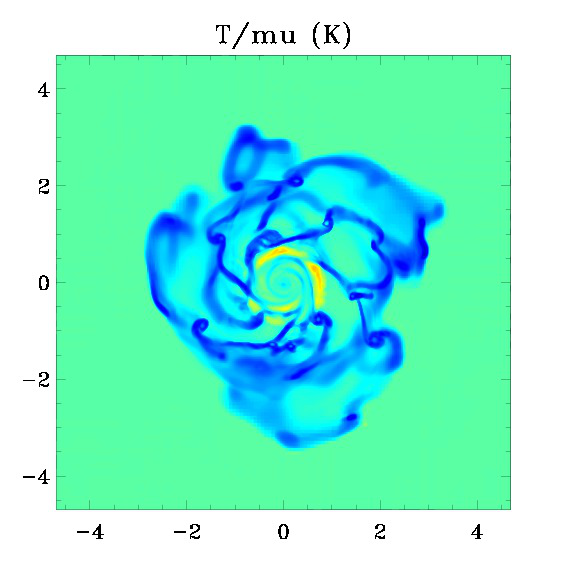}
    \includegraphics[width=0.3\textwidth]{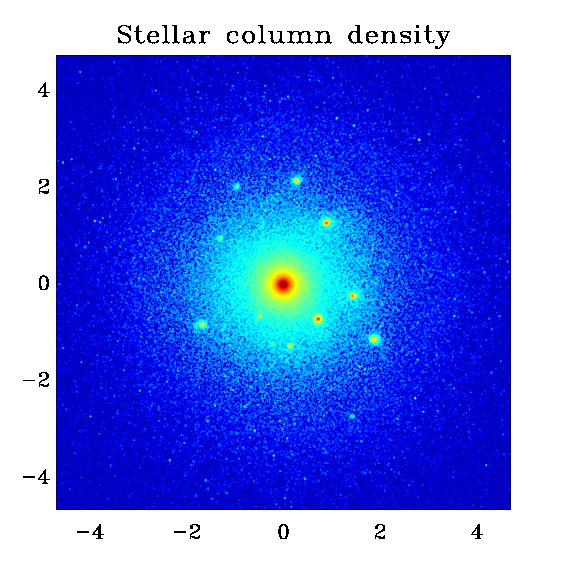}\\
    
 \caption{Maps of gas density (left, in logarithmic scale between 0.1 et 1000~H/cc), gas temperature (middle, in logarithmic scale between 300 and $10^6$~K) and stellar column density (right, logarithmic scale in arbitrary units), seen face-on (lower plot) and side-on (upper plot), with (upper half) and without (lower half) feedback after 2~Gyr of evolution. The images are all 10~kpc across.}
  \label{fig:maps}
\end{figure*}

The last but probably most important ingredient in galaxy formation simulations is stellar feedback. Stars, especially young stars, interact quite strongly with their environment. Among many possible physical processes, HII regions, stellar winds, type II  and type Ia supernovae explosions have been the most popular. These various mechanisms are believed to be responsible for SF regulation in dwarf galaxies \citep{1986ApJ...303...39D}, and they contribute collectively to roughly a couple of $10^{51}$~erg of energy injection into the surrounding ISM, for an average $10~M_{\odot}$ massive star and for a standard stellar initial mass function (IMF) \citep{2011MNRAS.410.2625P}. Since all these processes occur at very small, sub-pc scales, we have to rely on rather approximate numerical implementations. The simplest approach is clearly to inject directly thermal energy into each gas cell containing young stars. It turns out that feedback is very inefficient in that case because it occurs mostly in dense gas where cooling immediately radiates away the added thermal energy. It is usually considered that this strong cooling is a spurious numerical effect \citep{2009ApJ...695..292C}. But supernovae explosions, and more generally feedback from young stars, takes place in very dense environment (such as molecular clouds) so that strong radiative effects are likely to take place anyway. More recently, radiative pressure from young stars has been also proposed as a new possible mechanism \citep{2010ApJ...709..191M} to drive momentum into the interstellar medium (ISM). This approach is very promising since gas momentum does not suffer directly from gas cooling. It is however quite difficult to model since it requires detailed radiative transfer calculation of the infrared radiation reprocessed by dust. It was also shown recently that, for dwarf galaxies such as the ones studied here, supernovae feedback likely dominates over other sources of feedback energy \citep{2012MNRAS.tmp.2654H}~: radiative effects become inefficient due to the lower column densities. 

Physical processes associated to stellar feedback are probably even more complex than the ones quoted here so far. For example, we know from X-ray observations of supernovae remnants than they are strongly magnetized and turbulent. Moreover, cosmic rays acceleration occurs very rapidly and the energy density stored in the relativistic component is large enough to affect significantly the dynamics of the propagating shock wave \citep{2004A&A...413..189E, 2010A&A...509L..10F}. These non-thermal processes have much longer dissipation time-scales than the thermal component, so they can store feedback energy longer than the rapidly cooling gas and release it to the gaseous component more gradually. Modeling unresolved turbulence, magnetic field amplification and cosmic ray propagation and radiative losses in the framework of galaxy formation is at its infancy, although some recent progresses have been reported \citep{2006AN....327..469H, 2008A&A...481...33J, 2010MNRAS.405.1634S, 2012MNRAS.tmp.3026U}. We propose here a much simpler formalism that captures very crudely these various non-thermal processes and couple more efficiently the energy associated with stellar feedback to the gas component. 

The idea is to introduce a new variable for the energy density in these non-thermal components. We note this new variable $e_{\rm turb}$ for simplicity, although it doesn't have to be associated to turbulence only, but also to cosmic rays and magnetic fields. The specific turbulent energy $\epsilon_{\rm turb}$ is defined by $e_{\rm turb} = \rho \epsilon_{\rm turb}$. We model the time evolution of this non-thermal energy using the following simple equation 
\begin{equation}
\rho \frac{D \epsilon_{turb}}{Dt}= \dot E_{inj} - \frac{\rho \epsilon_{turb}}{t_{\rm diss}}
\end{equation}
where we have only 2 terms, the non-thermal energy source $\dot E_{inj}$ due to stellar feedback and the energy dissipation modeled classically as a damping term with dissipation time-scale $t_{\rm diss}$. The time scale depends on the exact underlying dissipation mechanism, such as radiative losses for cosmic rays \citep{2007A&A...473...41E} or eddy turn-over time for turbulence. In what follows, we just need to accumulate enough feedback energy to drive strong outflows and to regulate our SF efficiency across the galactic disk. In this paper, we used a fixed dissipation time scale
\begin{equation}
t_{diss} \simeq 10~{\rm Myr}
\end{equation}
comparable to the typical molecular cloud life time, as expected from their observed internal SF efficiency  \citep{1997ApJ...476..166W}.
Other options are possible, such as, for example, $t_{diss} \simeq \lambda_{\rm J} / \sigma_{\rm turb}$, where $\lambda_{\rm J}$ could be the typical Jeans length or, equivalently, the typical cell size, or even more exotic, the cosmic rays mean free path or magnetic dissipation length scale. The turbulence velocity dispersion $\sigma_{\rm turb}$ is defined here as
\begin{equation}
\epsilon_{turb}= \frac{1}{2}\sigma^2_{turb} {\rm .}
\end{equation}
We parametrized the non-thermal energy injection as 
\begin{equation}
\dot E_{inj} = \dot \rho_* \eta_{\rm SN}10^{50}~{\rm erg/M_\odot} {\rm ,}
\end{equation}
which corresponds to a maximally efficient model for which $10^{51}$~erg per $10~M_{\odot}$ massive star is injected into non-thermal energy. For a typical IMF, the mass fraction of stars going supernovae is roughly $\eta_{\rm SN}=0.1$. Although these numbers are determined by stellar evolution models, we believe that they can be considered as free parameters. Indeed, both the IMF and stellar evolution models are poorly constrained at low metallicity and/or at high redshift, and more generally in extreme galactic environments \citep{2012MNRAS.tmp.2706M}. Note that this energy is injected in the gas cell only 10~Myr after the star particle has been created. 

The final step of our feedback model is to compute the non-thermal pressure using $P_{\rm turb} = \rho \sigma_{\rm turb}^2$ and add it to the thermal pressure to get
\begin{equation}
P_{\rm tot} = P_{\rm thermal} + P_{\rm turb} {\rm .}
\end{equation}
This obviously requires to modify the Euler equation by adding the non-thermal pressure term, and this could impact more or less deeply the hydrodynamics solver. One even simpler alternative is to actually add directly to the thermal energy the injected non-thermal energy as
\begin{equation}
\rho \frac{D \epsilon_{\rm thermal}}{Dt}= \dot E_{\rm inj} - P_{\rm thermal}\nabla \cdot {\bf v} - n_H^2 \Lambda
\end{equation}
and shut down gas cooling to mimic the contribution of the non-thermal pressure to the total pressure. $\epsilon_{\rm turb}$ becomes a completely independent variable (like a passive Lagrangian tracer) used only to re-activate cooling when the non-thermal contribution becomes comparable or smaller than the thermal energy. In practice, we shut down cooling everywhere the turbulence velocity dispersion is large enough, using    
\begin{equation}
\Lambda=0 ~{\rm if}~\sigma_{\rm turb} > 10 ~{\rm km/s} {\rm .}
\end{equation}
Our final model does not require any modification to the hydro solver. It just requires to shut down cooling in each cell, if the velocity dispersion is above a chosen threshold (here 10~km/s). The other advantage is that it is qualitatively very similar to the ``delayed cooling'' approach used in other recent studies \citep{2006MNRAS.373.1074S, 2010Natur.463..203G, 2011MNRAS.410.1391A}, although our current approach is motivated by non-thermal astrophysical processes for which a detailed modeling would be far beyond the scope of the present study.

\section{Evolution of the dark matter density profile}

\begin{figure*}
    \includegraphics[width=0.45\textwidth]{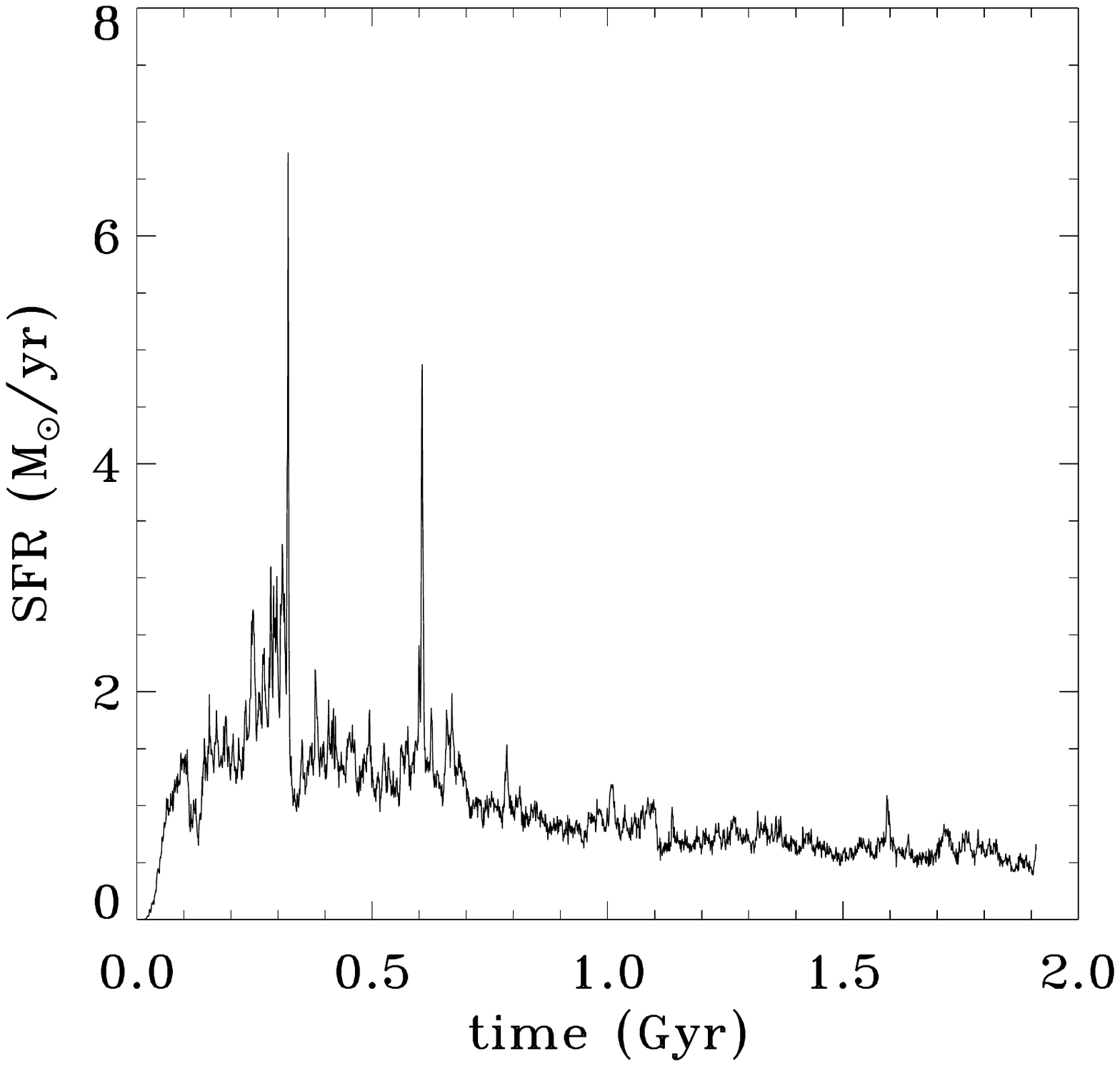}
    \includegraphics[width=0.45\textwidth]{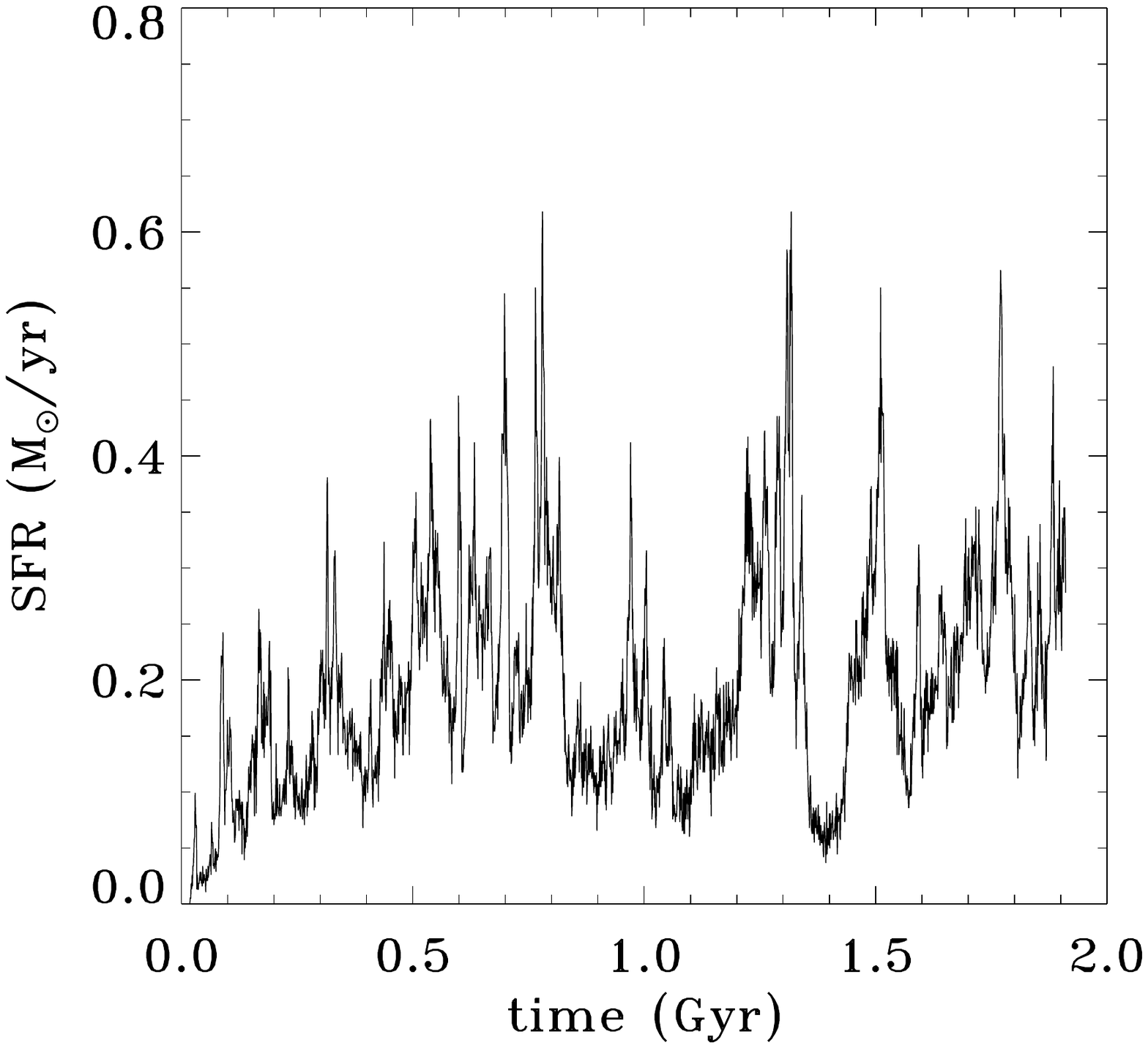}
 \caption{ Star formation history in the runs without (left plot) and with (right plot) feedback. }
  \label{fig:sfr}
\end{figure*}

\begin{figure}
    \includegraphics[width=0.5\textwidth]{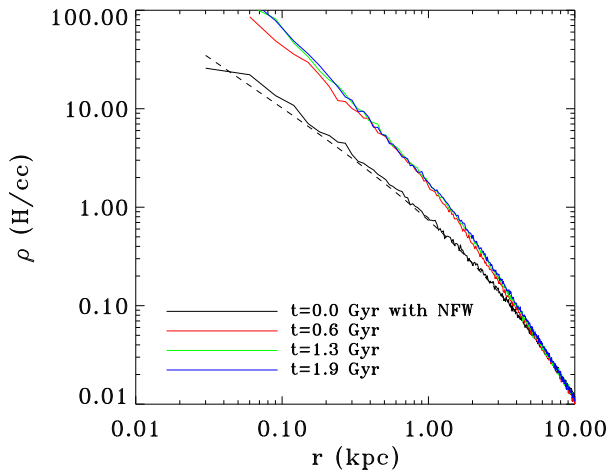}
  \caption{Evolution of the dark matter density profile over the 2 Gyr of evolution for the control run with cooling, star formation but no feedback.   The dark matter halo has been strongly adiabatically contracted. }
  \label{fig:nofbkrun}
\end{figure}

\begin{figure}
    \includegraphics[width=0.5\textwidth]{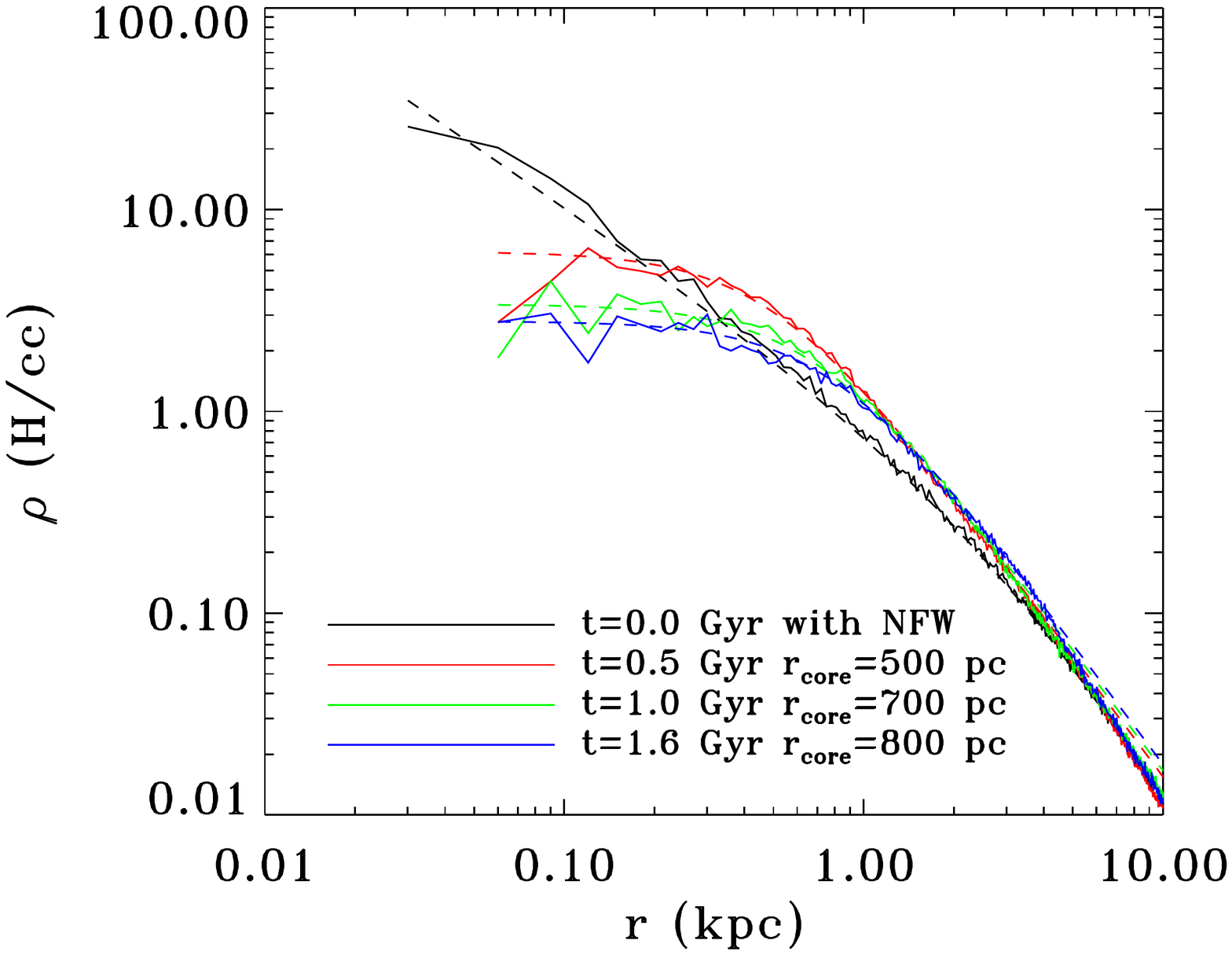}
  \caption{Evolution of the dark matter density profile over the 2 Gyr of evolution for the control run with cooling, star formation and stellar feedback. We see the formation of a large core. We also show for comparison the analytical fit (dashed line) based on a pseudo-isothermal profile (see text for details) }
  \label{fig:fbkrun}
\end{figure}

Now that we have presented our numerical set-up and parameters, we now describe in more details our results. We have performed 3 different simulations, in order to highlight the effect of stellar feedback on the mass distribution within the halo. The first simulation, already presented in the previous sections, was a pure adiabatic gas evolution of our initial hydrostatic equilibrium halo. We checked that our initial set-up was indeed stable over 2~Gyr time, as it should be since no gas cooling was considered in this case (see Fig.~\ref{fig:dmorun}). We would like to stress that this is a very important step in our methodology, since it demonstrates that any evolution in the dark matter density profile has to be related to the dissipative nature of baryons, through gas cooling, star formation or feedback. The second simulation was run with (metal dependent) gas cooling and star formation. No stellar feedback was included. In this case, the gas loses pressure support and rains down towards the centre of the halo, quickly reaching the centrifugal barrier and sets up into a centrifugally supported disc. As one can see in Figure~\ref{fig:maps}, the final disk is very thin and fragments into dense gas clumps that form stars actively. This gives rise to the formation of dense bound star clusters that survive for long times. We also see the formation of a massive bulge in the centre of the galaxy, leading to an overall highly concentrated baryons distribution. The associated SF history can be seen on Figure~\ref{fig:sfr}: it is on average very high, around 1~$M_{\odot}$/yr, with short bursts reaching 4 to 6~$M_{\odot}$/yr, associated to the formation of dense gas clumps. Such a high SF rate is usually associated to massive galaxies at low redshift. This is quite unrealistic for dwarf galaxies we see today \citep{2002AJ....124..862H}. The effect of this strongly dissipative evolution on the dark matter profile can be seen on Figure~\ref{fig:nofbkrun}. After 1~Gyr, the dark matter distribution has been adiabatically contracted very significantly by baryons. The inner slope of the dark matter density profile is close to -2, and no core is visible. It is worth mentioning that although we have a very clumpy structure in the ISM and in the stellar distribution, it does not trigger the formation of a dark matter core in our case: the mechanism proposed by  \cite{2006Natur.442..539M} \citep[see also][]{2001ApJ...560..636E} does not work here, probably because our clumps are not massive enough. 

We now move to our final run with gas cooling, SF and stellar feedback. The evolution of the star forming disk is dramatically different from the  ``no feedback'' run. We see in Figure~\ref{fig:maps} that the final gas distribution shows a very thick, turbulent disk, with strong outflows made of shredded clouds and filaments. The face-on view reveals that many gas clouds form in the outskirts of the disk, while the central region has been evacuated by stellar feedback, giving rise to the wind. The temperature map illustrates nicely the hot gas in the wind, segregating from the cold gas in the ISM. Star formation still proceeds within dense clouds, but these are not long-lived anymore. This is why we don't see any massive star clusters in the stellar surface density map. Only a few managed to survive. This is one of the key qualitative features of our stellar feedback implementation: gravitational instability and shock compression trigger the formation of star forming clouds, which are then quickly disrupted by stellar feedback, recycling the unused gas into the ISM and giving rise to the galactic wind. 

The SF rate plotted in Figure~\ref{fig:sfr} is one order of magnitude lower in average than the ``no feedback'' case. It exhibits strong bursts followed by quiescent phases. When gas cools down and sinks towards the central region, SF rises sharply and triggers a starburst. Stellar feedback then removes the gas into the hot wind, leaving the central kpc almost devoid of gas. This explains the very low star formation episodes. Gas then rains back down from the corona and triggers a new SF episode. These cycles are clearly visible in the SF history. They are at the origin of strong potential fluctuations due to massive periodic gas outflows and inflows.

In Figure~\ref{fig:fbkrun}, we plot the evolution of the dark matter density profile. As a technical side note, we would like to stress that these profiles have been computed using the highest dark matter density peak as centre, defined using the ``shrinking sphere'' algorithm. This minimizes spurious features in the density profiles due to poor centering. We clearly see in the ``feedback'' run that a large dark matter core develops. 
We tried to fit the resulting profile using a ``pseudo-isothermal'' profile defined as
\begin{equation}
\rho \propto \frac{1}{1+(r/r_{core})^2}
\end{equation}
This analytical shape has traditionally been used in rotation curve fitting and has proven to work extremely well in nearby dwarf galaxies \citep{1991MNRAS.249..523B, 2006ApJS..165..461K,2008ApJ...676..920K}. As can be seen in Figure~\ref{fig:fbkrun}, this is also true in our simulation, for which the pseudo-isothermal fit is remarkably good within the central 3-4~kpc. We have measured core radii ranging from 500~pc at early time (0.5~Gyr) to 800~pc at late time (1.6~Gyr). We do see a systematic increase of the core radius with time. We have also tried to fit our dark matter profile with other popular cored density distributions \citep[][for example]{1995ApJ...447L..25B} but with less success. 

To quantify even more the time evolution of the dark matter distribution, we have fitted the density profile between 200~pc and 800~pc with a single power law. This is obviously a poor fit to our cored distribution, but this captures the essence of the dark matter flattening in the centre.  
At $t=0$, we measure for the slope $\alpha=-1.2$, which is the average value of the NFW profile between the 2 chosen radii. At the beginning, gas cools down and adiabatic contraction of the halo can be measured as a decrease of the inner slope to $\alpha=-1.8$. But stellar feedback processes very quickly develop and trigger strong potential fluctuations, leading to a gradual flattening of the slope. We end up after 2~Gyr of evolution with a slope between $\alpha=-0.3$ and $\alpha=-0.5$, in excellent agreement with the measurement reported by PG12 in the context of a lower resolution cosmological simulation. 

As proposed by PG12, this strongly suggests that the core formation mechanism is related to violent gas outflows triggered by a succession of starbursts. In order to validate this idea in our simulation, we have computed the time evolution of the enclosed gas mass within spheres of increasingly large radii, from 200 pc for the smallest to 1600 pc for the largest. The results are shown in Figure 7 and illustrate that gas is repeatedly removed from the central region of the galaxy, giving rise to strong coherent potential fluctuations. As more gas cools down from the halo, the SF bursts become more violent and large mass fluctuations propagate to larger radii. This explains why we see a systematic increase in the dark matter core size as a function of time. At the end of the simulation, only scales up to $\sim$800~pc are affected by these strong fluctuations, corroborating nicely the measured core size of 800 pc.

While these results are qualitatively consistent with the analysis of PG12, there is a difference in the quantitative details.  In our case, many potential fluctuations are not only established in under a dynamical time (as in PG12) but also are erased before a full dynamical time has elapsed (unlike in PG12). The effect of these short-time transients is ignored by the PG12 toy model, since it starts by sampling the potential only once per dynamical time. Applying the PG12 algorithm directly therefore leads to incorrect predictions for our present simulation. We verified, however, that consistent predictions can be made by changing the sampling method to pick out, at each dynamical time step, the most extreme potential jump that occurred within the simulation. While this is a heuristic method, we believe it shows that the essential ingredient of impulsive potential changes is at play in our new simulations. It is in principle possible to extend the PG12 work to track phases of particles, thus giving a complete description of the impulsive changes which does not require the problematic quantization of time. However this approach would lose the attractive simplicity of the PG12 model and is certainly beyond the scope of the present investigation.

We should also emphasis that the core forms after only 0.5~Gyr in our reference feedback run, when only 10\% of the stars have formed. We have indeed measured a total stellar mass of $M_*\simeq 5\times 10^7$~M$_\odot$ at that time. This means that the final total energy liberated in supernovae explosions is not the limiting factor. This is in fact consistent with the estimates of energy required to form cores in \cite{2012arXiv1207.2772P}. The question of what limits star formation requires full cosmological runs to tackle further, since our "monolithic collapse" models do not reflect the interplay between gas inflow, outflow and feedback that presumably regulate star formation as of function of redshift in the real universe.
  
\begin{figure}
    \includegraphics[width=0.5\textwidth]{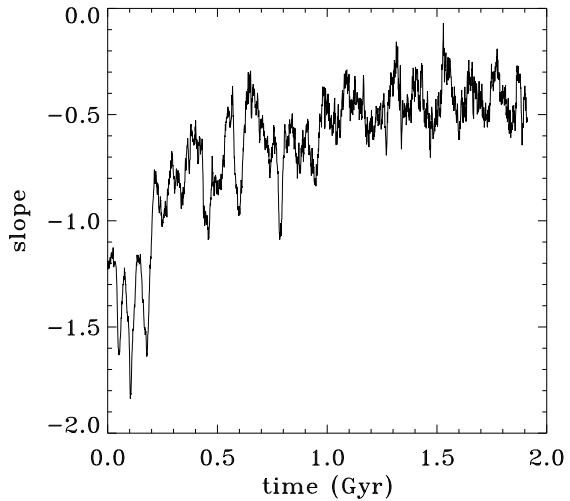}
  \caption{Time evolution of the slope of the dark matter density profile measured between 200 and 800 pc for the simulation with feedback.  }
  \label{fig:slope}
\end{figure}

\begin{figure}
    \includegraphics[width=0.5\textwidth]{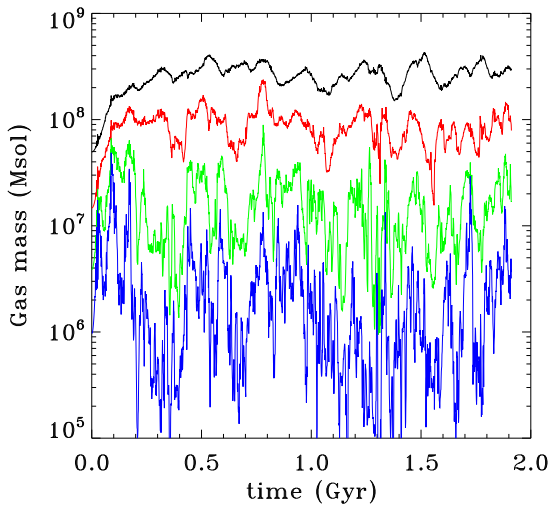}
 \caption{ Time evolution of the total enclosed gas mass within spheres of radius 200 (blue), 400 (green), 800 (red) and 1600 (black) pc for the simulation with feedback. }
  \label{fig:gasmass}
\end{figure}

\section{Observational galaxy properties}
\label{sec:galprop}

We have shown that cusp-core transformations are expected if there are strong (order unity) potential fluctuations on dynamical timescales. Such fluctuations are non-adiabatic, non-reversible, and heat both the dark matter and stars. We discuss here the observational consequences of such fluctuations; these can then be used to constrain or rule out the presence of such fluctuations in real systems. 

\subsection{The stellar distribution} 

\begin{figure}
    \includegraphics[width=0.5\textwidth]{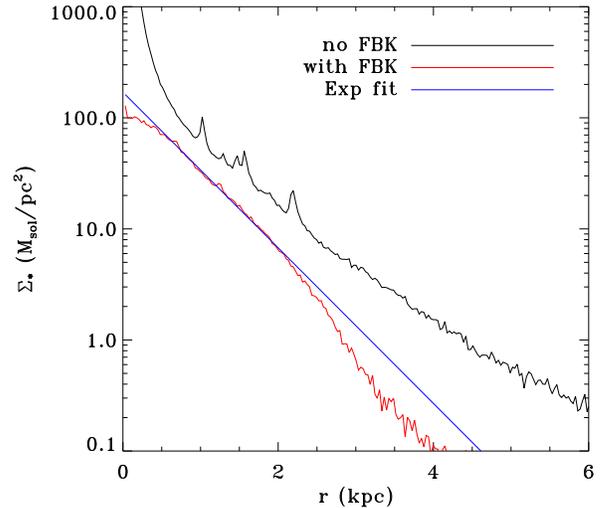}
 \caption{Stellar surface density of the final dwarf galaxy with (red line) and without (black line) feedback. In blue is shown the exponential fit of the stellar disk with feedback with scale length $R_d = 1.1$~kpc. }
  \label{fig:surfdens}
\end{figure}

\begin{figure}
    \includegraphics[width=0.5\textwidth]{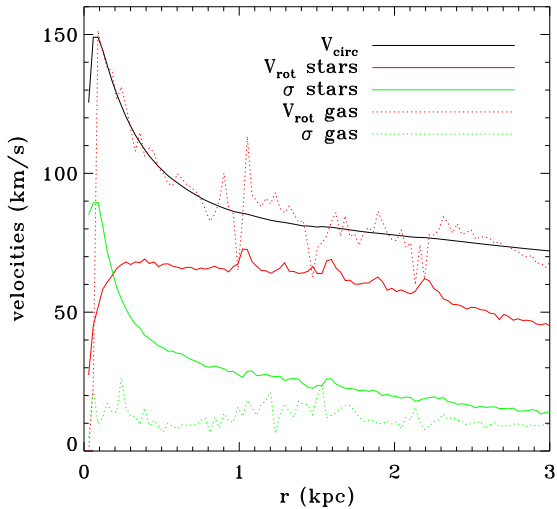}
 \caption{Kinematic analysis of our dwarf galaxy without feedback: circular velocities (black solid line), stellar tangential velocity (red solid line) and stellar tangential velocity dispersion (green solid line), compared to the gas tangential velocity (red dotted line) and gas tangential velocity dispersion (green dotted line). }
  \label{fig:nofbkkin}
\end{figure}

\begin{figure}
    \includegraphics[width=0.5\textwidth]{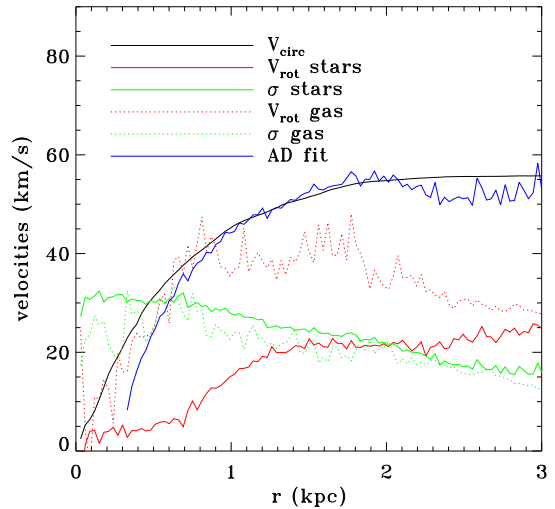}
 \caption{Kinematic analysis in the feedback case: circular velocities (black solid line), stellar tangential velocity (red solid line) and stellar tangential velocity dispersion (green solid line), compared to the gas tangential velocity (red dotted line) and gas tangential velocity dispersion (green dotted line). Also shown as the blue solid line is the predicted circular velocity curve based on the Asymmetric Drift (AD) approximation.}
  \label{fig:fbkkin}
\end{figure}

\begin{figure}
    \includegraphics[width=0.5\textwidth]{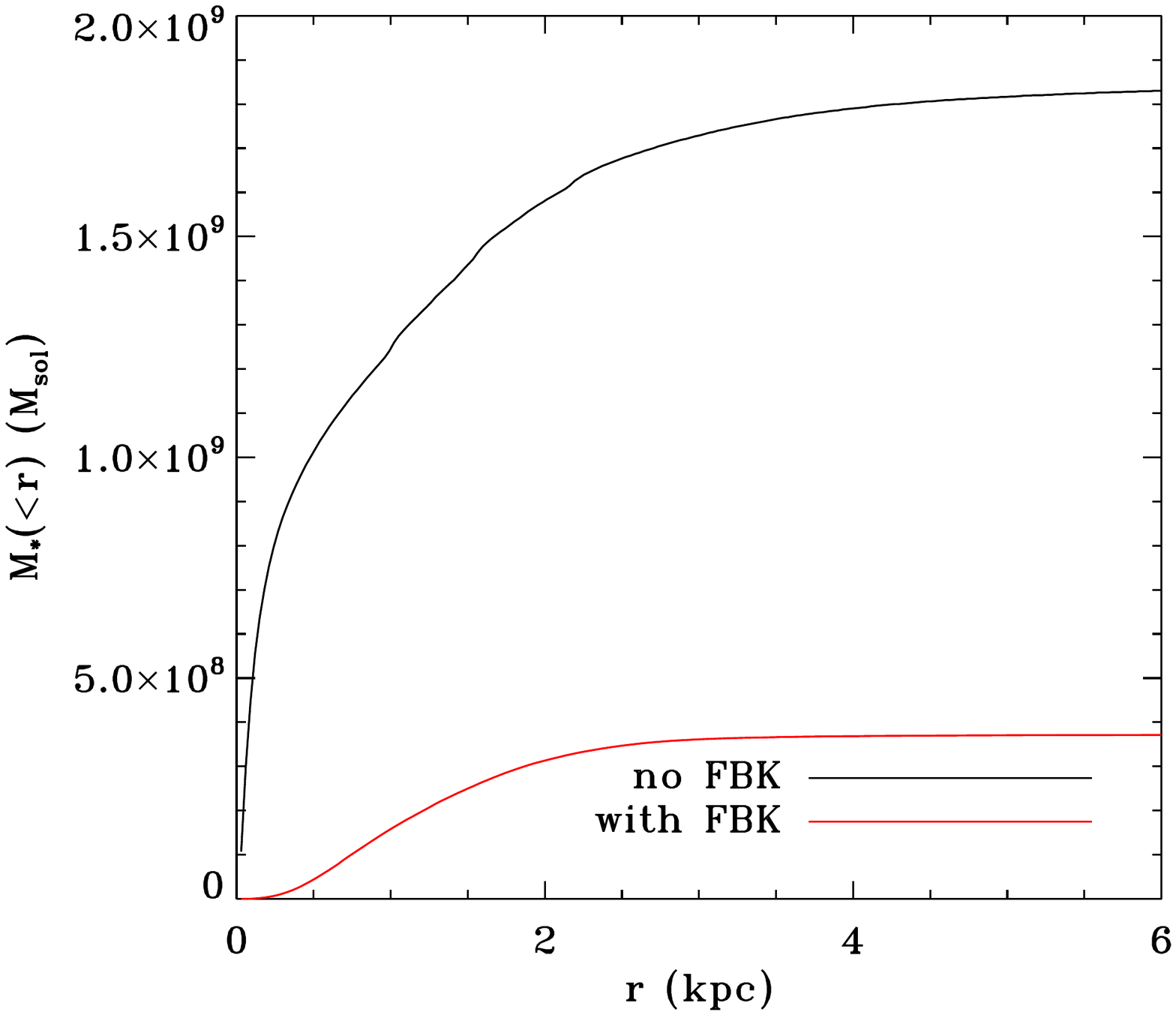}
 \caption{Cumulative stellar mass profile as a function of spherical radius for the fiducial star formation efficiency (1\%) without (black line) and with (red line) stellar feedback at $t \simeq 1$~Gyr. Also shown are the stellar mass profiles with a very low star formation efficiency (0.2\%) and a top heavy initial mass function, without (green line) and with (blue line) stellar feedback. Only the latter model can match the observed stellar mass in the dwarf galaxy WLM.}
  \label{fig:mstar}
\end{figure}

\citet{2005MNRAS.356..107R} found from their toy models that, when cusp-core transformations occur, the resulting surface brightness of the stars is well-approximated by an exponential over many scale lengths, sometimes with a break radius at large radii and a core at small radii. 
This is exactly what we see in the present simulation: in Figure~\ref{fig:surfdens}, we have plotted the stellar surface density with and without feedback. In the latter case, we see a prominent bulge with high Sersic index and a weak exponential disk extending up to 6~kpc, while in the feedback case, we see a smaller exponential disk, with no bulge in the center and a clear signature of a stellar core within the central 500~pc. 
We have fitted this stellar distribution with a pure exponential disk with scale length $R_d=1.1$~kpc (see Figure~\ref{fig:surfdens}).   
In our simulation without feedback,  the stars form in a razor-thin, rotationally supported disc. 

Note that simulations of the formation and evolution of isolated dwarf galaxies generically end up with near-exponential surface brightness profiles, with rather thick gas and stellar disks \citep{2003MNRAS.339..289S,2008MNRAS.389.1111V,2008A&A...477...79D}. These simulations can be however divided into 2 main categories: quiescent feedback models, in which small scale effects are captured by an effective equation of state, leading to a quiescent star formation history where outflows are obtained through a quasi-stationary galactic wind \citep[see e.g.][]{2003MNRAS.339..289S,2008A&A...477...79D}; and bursty models, in which the star formation history is not stationary and shows violent fluctuations \citep[see e.g.][]{2007ApJ...667..170S,2008MNRAS.389.1111V,2009A&A...501..189R}. The first category of simulations give rise to moderately thick disk, and very weak potential fluctuations. The stellar disk is still well defined with an aspect ratio close to $h/R \simeq 0.1$ \citep[see Fig. 11 in ][for example]{2003MNRAS.339..289S}. Therefore, we should not expect any dark matter core in this case. The second type of simulations, however, gives much thicker disks, in some case almost spheroidal galaxies. They show violently time-varying outflows, and they should lead to the formation of a dark matter core. Our present model, with strong stellar feedback, belongs to the second category: the resulting stellar distribution is oblate and quite hot, with $v/\sigma \simeq 1$ and a scale height to scale length ratio close to 0.5 (see Figure~\ref{fig:maps} and Figure~\ref{fig:fbkkin}).

Unfortunately, it can be difficult to conclusively test the above expectations. Exponential surface brightness profiles can form through a variety of physical processes not related to violent potential fluctuations \citep[e.g.][]{1987ApJ...320L..87L}. Similarly, hot stellar distributions with $v/\sigma \sim 1$ need not be a smoking gun for cusp/core transformations. Galactic mergers \citep[e.g.][]{1978ApJ...225..357S,2006MNRAS.371..885R}, and collisionless heating from a strong tidal field \citep[e.g.][]{2001ApJ...559..754M,2011ApJ...739...46L} both lead to stellar distributions with low $v/\sigma$. 

We can avoid all of the above complications, however, if we focus on {\it isolated} low mass dwarf galaxies. The best-studied system to date is WLM. Despite being extremely tidally isolated, WLM does indeed have a hot, oblate spheroidal, stellar distribution (with $v/\sigma \simeq 1$ and $h/R_d \simeq 0.5$) that is reasonably approximated by an exponential surface brightness fall-off \citep{2012ApJ...750...33L}. Note that in this case too, a flattening on the stellar surface density profile is seen in the centre, reminiscent of a stellar core\footnote{PG12 pointed out the heating processes applying to the DM should also apply to the stars.}.
\cite{2012ApJ...750...33L} point out that it may be difficult to understand why the stellar distribution in WLM is so hot without recourse to strong and bursty stellar feedback. We can confirm that view here. In our model of an isolated dwarf galaxy, only the simulation with very strong feedback managed to form a hot thick-disc-like stellar distribution. This same simulation found a significant cusp-core transformation in the underlying dark matter distribution. In order to push the comparison with WLM even further, we have analyzed the kinematic properties of our dwarf galaxy. These are shown on Figure~\ref{fig:fbkkin}. One can see immediately that the stellar rotation curve is slowly rising, while the stellar velocity dispersion is slowly declining. Both curves intersect at around 2~kpc, at a value close to 20~km/s. These features are in striking agreement with WLM data as exposed in \cite{2012ApJ...750...33L}. The gas kinematic properties are also very similar, as shown in Figure~\ref{fig:fbkkin}.  Using our kinematic analysis, we are in a good position to test one important method to derive the total mass profile in dwarf galaxies from their kinematic properties, namely the Asymmetric Drift (AD) model. This method, based on the Jeans equation, follows a few reasonable simplifying assumptions to derive the relation between the circular velocity and the velocity moments. Following \cite{2001AJ....121..683H} and \cite{2012ApJ...750...33L}, we used $v^2_{\rm circ} = v^2_{\rm \theta} + \sigma^2_{\rm \theta} \left( 2r/R_d-1\right)$ with $R_d \simeq 1.1$~kpc, as measured in our simulation. We see in Figure~\ref{fig:fbkkin} that AD is overall a good approximation to recover the underlying mass profile, except perhaps in the very center where it is underestimated. The total mass inferred from this analysis by \cite{2012ApJ...750...33L} for WLM, $M_{\rm tot} \simeq 2\times 10^{10}$~M$_{\odot}$ is therefore accurate, and again very close to our simulated halo mass. 

Although our spatial and kinematic properties are in striking agreement with the relatively isolated dwarf WLM, the total stellar mass that we obtained in our simulation is too large by one order of magnitude. We have plotted in Figure~\ref{fig:mstar} the cumulative stellar mass profile in spherical shells. One sees clearly that without feedback, almost all baryons are converted into stars after 1~Gyr, since we get $M_{\rm *} \simeq 10^9$~M$_{\odot}$. With our strong stellar feedback model, we managed to reduce this number by one order of magnitude, down to $M_{\rm *} \simeq 10^8$~M$_{\odot}$. This is quite an achievement, but it falls short by one order of magnitude to explain the stellar mass observed in WLM, which has been measured by \cite{2007ApJ...656..818J} to be  $M_{\rm *} \simeq 1.1\times 10^7$~M$_{\odot}$. We are therefore overproducing stars to a level comparable to most current galaxy formation simulations \citep{ 2009arXiv0909.4167P, 2010Natur.463..203G, 2011MNRAS.410.1391A}, when compared to individual galaxies or to an ensemble of galaxies using the abundance matching technique \citep{2010MNRAS.404.1111G, 2010ApJ...710..903M}, although recently \cite{2012arXiv1209.1389M} argue differently.  Although solving this issue is beyond the scope of the present paper, we have a conceptually simple way to solve this problem, by lowering the star formation efficiency parameter $\epsilon_*$ by one order of magnitude, and in the same time, increasing the mass fraction of massive star going supernovae by also one order of magnitude. The first idea could be justified by the low metallicity we find in dwarf galaxies, leading to a inefficient regime of star formation, for which dust shielding is less efficient at promoting $H_2$ molecule formation \citep{2011arXiv1106.0301K}. The second idea could be justified by recent observations of low metallicity star clusters in the Galaxy, which are consistent with a top-heavy IMF \citep{2012MNRAS.tmp.2706M}. Using these two non-standard but plausible ingredients, we will straightforwardly obtain the same energy input from supernovae, and therefore the same hydrodynamical model, with however ten times less {\it long-lived} star formed. A different model based on early radiative stellar feedback has been recently proposed by \cite{2012arXiv1201.3359B} that could solve this very same issue, although the quantitative effect would not be as straightforward as the one proposed here. 

Since complex non-linear hydrodynamical processes are at play, it is not obvious that lowering $\epsilon_*$ (while increasing the feedback energy per stellar mass formed) will result in decreasing the total stellar mass formed. We therefore run 2 additional simulations with $\epsilon_*=0.2\%$ and a top-heavy IMF so that 50\% of the mass of a given stellar population (or star particle) will be in massive stars going supernovae. The first simulation was run without stellar feedback (only metal enrichment) and the second simulation was run with our new stellar feedback scheme and a top-heavy IMF. We have plotted in Figure~\ref{fig:mstar} the cumulative stellar mass profiles of these 2 additional runs. Interestingly, although we have decreased by a factor of 5 the SFE, the run without feedback gives also rise to the formation of  $M_{\rm *} \simeq 10^9$~M$_{\odot}$ of stars, like in the fiducial case. This means that, without stellar feedback, the gaseous disk always manages to cool and contract enough to reach high gas densities and transform most of its baryons into stars. With stellar feedback, however, we form only $M_{\rm *} \simeq 2\times10^7$~M$_{\odot}$ of stars after 1~Gyr, in much better agreement with the observed stellar mass in WLM. We have also checked (not shown here) that, in this last simulation, all of the galaxy properties discussed above are also recovered, namely a large dark matter core, thick stellar and gaseous disks and a bursty SF history, with large amount of gas moving in and out of the central kpc of the galaxy. This last model gives very encouraging results, but requires a significantly top-heavy IMF together with a very low star formation efficiency. We would like also to stress again (see Section~4) that star formation might be also regulated or even stopped after only 0.5~Gyr of evolution, due to the effect of a more realistic cosmological accretion history, which is not captured here in this idealized set-up.

\subsection{Star formation histories} 

A second key observational prediction of cusp-core transformations is the star formation history (SFH). In the category of feedback models where cusp-core transformations are expected \citep[][and this paper]{2007ApJ...667..170S,2008MNRAS.389.1111V,2009A&A...501..189R,2012MNRAS.423..735C}, the SFH is extremely bursty with peak to trough variations of 10 and a duty cycle of roughly one dynamical time. Although our {\it average} star formation rate of 0.02 to 0.2~$M_\odot/$yr for a total gas mass of $10^9~M_{\odot}$ is consistent with observations of blue compact dwarf galaxies in the local universe \citep{2002AJ....124..862H}, the bursty nature of its time evolution needs to be tested observationally. There are two possibilities: (i) measure the SFH for individual systems; and (ii) estimate the {\it variance} in the star formation statistically for a large population of like-galaxies. The former is cleaner, but requires very high time resolution in the derived SFH; the latter has potentially high time resolution but relies on assumptions about the equivalence of large populations of galaxies. We discuss each in turn, next. 

\subsubsection{Individual star formation histories} 

The best SFHs come from the most nearby systems. These have resolved color magnitude diagrams (CMDs) that reach the oldest main sequence turn-offs, which is vital for correctly recovering the intermediate age stars \citep[e.g.][]{2009ApJ...705.1260N}. More distant systems have SFHs derived via spectral energy distribution (SED) fitting of the unresolved stars that is degenerate and less accurate than CMD fitting \citep[e.g.][]{2012AJ....143...47Z}.

The best-studied system to date is the nearby dwarf spheroidal galaxy Sculptor that orbits the Milky Way \citep[e.g.][]{1998ARA&A..36..435M,2010MNRAS.406.2312L}. Unfortunately, even for excellent data, CMD fitting has a temporal resolution poorer than $\sim$1\,Gyr \citep{2002MNRAS.332...91D}, where the errors come from a mix of photometric uncertainty, spread in distance modulus, and errors in the stellar population evolution libraries \citep{2012ApJ...751...60D}. \citet{2012A&A...539A.103D} have recently attempted to improve this situation by simultaneously fitting the CMD and spectroscopic metallicities of red giant branch stars. They explicitly consider how well their methodology could recover a bursty star formation history. Using simulated data, they show that their method would recover a smooth continuous star formation history (as observed for Sculptor) from an input bursty history. Thus, the very best data remain inconclusive. It is interesting to note, however, that the very latest dynamical models for Sculptor appear to favor a cored rather than cusped mass distribution \citep{2008ApJ...681L..13B,2011ApJ...742...20W,2012MNRAS.419..184A}. This motivates a continued effort to attempt to confirm or deny a highly bursty star formation history for Sculptor.

\subsubsection{The star formation history of a population of dwarf galaxies}

An alternative approach to studying individual galaxies is to study a population statistically. If we assume that all dwarf galaxies of a given mass and type are statistically equivalent, then we can treat these has individual sample points along the SFH. If the SFH is smooth and continuous then the variance in the measured SFH for the population will be small; if, however, it is bursty then the variance will be larger. \citet{2012ApJ...744...44W} have recently used this idea to test a simple bursty star formation history against a population of 185 galaxies from the {\it Spitzer} Local Volume Legacy survey. They find that their more massive galaxies ($M_* \ge 10^7$\,M$_\odot$) are better fit (on average) by a smooth and continuous SFH, while the lower mass systems ($M_* \le 10^7$M\,$_\odot$) favour a bursty SFH with bursts of $\sim 10$'s of Myrs, inter-burst periods of $\sim 250$\,Myrs, and peak to trough burst amplitude ratios of $\sim 30$. These numbers are in total agreement with the SFH we obtained for our model with a cusp-core transformation (see Figure~\ref{fig:sfr}). WLM is again interesting here. It has a stellar mass of $1.1 \times 10^7$\,M$_\odot$ and a dynamical mass of $\sim 10^{10}$\,M$_\odot$ \citep{2012ApJ...750...33L}. As we have shown, it is comparable to our simulated halo, and statistically in the stellar mass range where bursty star formation in dwarf galaxies is observationally favored.

It seems that all of the currently available data favor a bursty SFH for dwarf galaxies with stellar mass $\le 10^7$\,M$_\odot$ and dynamical mass $\le 10^{10}$\,M$_\odot$. As we have shown here, such a bursty SFH will naturally heat the stars producing a hot stellar distribution even in isolated systems, consistent with recent observations of the isolated dwarf galaxy WLM. It also drives cusp-core transformations in the dark matter that can explain the now long-standing cusp-core problem \citep{1994ApJ...427L...1F,1994Natur.370..629M}. 

\section{Discussion}
\label{sec:summary}

In this paper, we have simulated an isolated dwarf galaxy using the now traditional set-up of the cooling halo. Using a new implementation of stellar feedback within the RAMSES code, we have modeled the formation of a dwarf galaxy of $10^{10}$~$M_{\odot}$ halo mass, whose evolution has been shown to be strongly dominated by feedback processes. We have observed the formation of a 800~pc dark matter core, the size of which corresponds roughly to 40 resolution elements. We have analyzed in details the stellar distribution and kinematics of our model galaxy, observable properties that compare successfully to the local isolated dwarf WLM. In our fiducial model, the total stellar mass is too large by one order of magnitude, suggesting that we need to lower the long-lived star formation efficiency, while keeping the same overall feedback efficiency to match all constraints. We have shown that this might be indeed achieved, if one uses a very low value for the Schmidt law parameter $\epsilon_* \simeq 0.2\%$, together with a top-heavy initial mass function.

Our simulation with feedback and AMR clearly confirms previous works performed with various SPH codes, both for isolated haloes and in a cosmological context, at a somewhat lower resolution than the one used here. The temporal variations and systematic evolution of the inner slope of the dark matter profile correlate nicely with bursts seen in the SF history of our dwarf galaxy.  We noted that there are several observational diagnostics which point to the cusp-flattening process driven by bursty star formation. The most obvious place to look is in star formation histories. Unfortunately for individual cases, it is unlikely to be possible to achieve sufficient time resolution \citep[e.g.][]{2012ApJ...751...60D, 2012A&A...539A.103D}. On the other hand, statistical studies of populations encouragingly point to bursts of exactly the right nature for cusp-flattening, and in agreement with our work, especially at low masses \citep{2012ApJ...744...44W}. 
There may be less direct evidence of a cusp-flattening process in the final kinematics of the stars. Since stars behave as collisionless particles, like dark matter, they should be heated by the irreversible process proposed by \cite{2005MNRAS.356..107R}, \cite{2006Natur.442..539M} and \cite{2012MNRAS.421.3464P}. Indeed when cusp-flattening occurs, we also obtain hotter, more oblate stellar distributions with $v/\sigma \simeq 1$. More directly related to the proposed mechanism, we also obtain a core in the stellar distribution. Promisingly, this also agrees well with the recent observations of the isolated dwarf galaxy WLM \citep{2012ApJ...750...33L}. 

Although these observations tend to favor bursty SF histories as a plausible origin to cusp-core transformations, we would like to stress that this is only a necessary condition for these to occur, not a sufficient one: other physical mechanisms could be at play to explain cusp-core transformations, not related to baryonic processes, but, for example, to warm or self-interacting dark matter particles.

\section*{Acknowledgments}
We thank our anonymous referee for helpful suggestions that greatly improved the quality of the paper.
We thank Pascale Jablonka for interesting discussions on the star formation history of dwarf galaxies.
Justin Read would like to acknowledge support from SNF grant PP00P2\_128540/1. 
The simulations presented here were performed on the COAST cluster at IRFU, CEA Saclay.


\bibliography{romain}


\label{lastpage}
\end{document}